\documentclass[12pt]{article}
\usepackage[centertags]{amsmath}
\usepackage{amsfonts}
\usepackage{amssymb}
\usepackage{amsthm} 
\usepackage{newlfont}
\usepackage{epsfig}
\usepackage{amscd}
\usepackage{color}

\newcommand{\beq}{\begin{equation}}
\newcommand{\eeq}{\end{equation}}
\newcommand{\ba}{\begin{array}}
\newcommand{\ea}{\end{array}}
\newcommand{\bea}{\begin{eqnarray}}
\newcommand{\eea}{\end{eqnarray}}

\begin{document}

\begin{center}
{\large \sc \bf Multidimensional linearizable system of $n$-wave type equations.
}

\vskip 15pt

{\large  A. I. Zenchuk }

\vskip 8pt

\smallskip

{\it  Institute of Problems of Chemical Physics, RAS,
Acad. Semenov av., 1
Chernogolovka,
Moscow region
142432,
Russia}

\smallskip

\vskip 5pt


\vskip 5pt

{\today}

\end{center}

\begin{abstract}
A linearizable version of multidimensional system of  $n$-wave type nonlinear PDEs is proposed.
This system is 
derived using the spectral representation of its solution via the procedure similar 
to the dressing method for the ISTM-integrable nonlinear PDEs. The proposed system is shown to 
be completely integrable, particular solution is represented.
\end{abstract}

\section{Introduction}

The well known (2+1)-dimensional $n$-wave equation \cite{K1,K2} integrable by the inverse scattering transform method (ISTM) 
\cite{GGKM,ZMNP,AC,Sh,ZSh1,ZSh2,ZM,K}
\begin{eqnarray}\label{Nw}
[C^{(3)},v_{t_2}]-[C^{(2)},v_{t_3}] + C^{(2)} v_{t_1} C^{(3)} - C^{(3)} v_{t_1} C^{(2)} +
[[C^{(3)},v],[C^{(2)},v]]=0
\end{eqnarray}
($C^{(i)}$ are constant diagonal matrices) gets a wide applicability in mathematical physics. 
In particular, it appears  in multiple-scale expansions 
of various physical systems, where it  describes the evolution of wave packets. 
{Eq.(\ref{Nw}) is the off-diagonal $N\times N$ ($N\ge 3$)  
first order partial differential equation (PDE) with 
 quadratic nonlinearity of special form and $(N-1)N/2\ge 3$ different directional derivatives 
 $\sum_{i=1}^3 V^{(\alpha\beta)}_i \partial_{t_i}$ along the vectors $V^{(\alpha\beta)}$
 with the coordinates 
 \begin{eqnarray}
 V^{(\alpha\beta)}_1= C^{(2)}_\alpha  C^{(3)}_\beta - C^{(3)}_\alpha C^{(2)}_\beta,\;\;
V^{(\alpha\beta)}_2= C^{(3)}_\alpha - C^{(3)}_\beta, \; \;V^{(\alpha\beta)}_3= C^{(2)}_\beta - C^{(2)}_\alpha. 
 \end{eqnarray}}
However, its natural higher dimensional generalization is not integrable. An example of (partially) 
integrable multidimensional generalization of eq. (\ref{Nw})  has a
significantly deformed structure. One of its version possessing the periodic lump 
solution is found in \cite{Z_JMP_2014}.

The nonlinear PDE discussed in this paper reads (written in components)
\begin{eqnarray}\label{Nwave}
&&
\sum_{m=1}^3 Z^{(m)}_{\alpha\beta}(v^{(lj)}_{\alpha\beta})_{t_m}+
\sum_{n=1}^2 T^{(n)}_{\alpha\beta}(v^{(lj)}_{\alpha\beta})_{x_n} - \sum_{k=1}^2
\sum_{\gamma=1}^N a^{(k)}_{\alpha\gamma\beta} (v^{(lk)}_{\alpha\gamma} v^{(0j)}_{\gamma\beta} -
 v^{(l0)}_{\alpha\gamma} v^{(kj)}_{\gamma\beta}) =0,\\\nonumber
&&
l,j=0,1,2,\;\;\alpha,\beta=1,\dots,N.
\end{eqnarray}
Here $v^{(lj)}$ are $N\times N$ matrix fields and $Z^{(m)}_{\alpha\beta}$, $T^{(n)}_{\alpha\beta}$, 
$a^{(k)}_{\alpha\gamma\beta}$ are constant coefficients  having a special  structure shown below in 
eqs.(\ref{c1}-\ref{c3}). 
The structure of this equation differs from that of (2+1)-dimensional $n$-wave equation (\ref{Nw}), 
in particular, it involves {
$6$ matrix fields  $v^{(lj)}$ instead of the single one.
Similar to eq.(\ref{Nw}), eq.(\ref{Nwave})  has the quadratic nonlinearity and $N(N-1)/2\ge 3$ directional derivatives 
$\sum_{m=1}^3 Z^{(m)}_{\alpha\beta} \partial_{t_i} + 
\sum_{n=1}^2 T^{(n)}_{\alpha\beta}\partial_{x_n}$. Since we have the 
five-dimensional space of independent variables, 
there are five independent directional derivatives.}

Although the formal identification of different $v^{(ij)}$ (more exactly, the reduction $v^{(ij)} =
C^{(i)} v^{(00)} C^{(j)}$ with some constant diagonal matrices $C^{(i)}$, $i=1,2$) is admittable by the structure of 
eq.(\ref{Nwave}), such the  reduction cuts the solution space drastically decreasing  its dimensionality,
so that the reduced system  loses its complete integrability. 

The feature of  system (\ref{Nwave}) is that  its linear limit ($v^{(ij)}\to w^{(ij)}$)
\begin{eqnarray}\label{Nwave_lin}
\sum_{m=1}^3 Z^{(m)}_{\alpha\beta}(w^{(ij)}_{\alpha\beta})_{t_m}+
\sum_{n=1}^2 T^{(n)}_{\alpha\beta}(w^{(ij)}_{\alpha\beta})_{x_n}  =0,\\\nonumber
i,j=0,1,2,\;\;\alpha,\beta=1,\dots,N,
\end{eqnarray}
is the same for all $i,j$.
{ Since eq. (\ref{Nwave})} is linearizable \cite{Calogero,CX1,CX2} its solutions are formally  related to the 
 solutions 
of  the associated system of linear partial differential equations  (PDEs) (\ref{Nwave_lin}). 
Deriving this relation, we turn to 
the spectral representation of solution as an auxiliary tool, and find a spectral function $\Psi(\lambda,\mu)$ of two spectral parameters. 
We call the 
algorithm deriving $v^{(ij)}$ in terms of the associated spectral function $\Psi$ a dressing algorithm in analogy with the dressing 
algorithm for equations integrable by ISTM \cite{Sh,ZSh1,ZSh2,ZM}.

Thus, the paper has the following structure. In Sec.\ref{Section:dressing} we represent a particular  dressing 
method which will be used for  deriving  the system of nonlinear PDEs (\ref{Nwave}). The supplemented 
algebraic-differential relations  among the fields of this system 
are derived in Sec.\ref{Section:relations}. Solution space  to system (\ref{Nwave}) is 
discussed in Sec.\ref{Section:solutions}, where an example of  particular solution is given.
Basic results are discussed in Sec.\ref{Section:conclusion}.

\section{Dressing method}
\label{Section:dressing}
\subsection{Derivation of spectral equation}
We consider the $N\times N$ matrix function $R(x;\lambda,\mu)$ depending on the
two spectral parameters $\lambda$ and $\mu$ and
set of auxiliary parameters $x=(x_1,x_2,\dots)$, which will be 
the independent variables of the nonlinear PDEs:
\begin{eqnarray}\label{R}
R(\lambda,\mu;x) = e^{ \sum_{j=1}^D s^{(j)}(\lambda) x_j +\sum_{k} S^{(k)}(\lambda) t_k } 
R_0(\lambda,\mu)  e^{ -\sum_{j=1}^D  s^{(j)}(\mu) x_j - \sum_{k} S^{(k)}(\mu) t_k} + r(\lambda) q(\mu),
\end{eqnarray}
where $r(\lambda) $ and $q(\mu)$ are some arbitrary diagonal functions of  spectral parameters, 
$s^{(i)}$  are independent arbitrary diagonal functions of  spectral parameter, while $S^{(m)}$ are linear 
combinations of $s^{(i)}$,
\begin{eqnarray}
S^{(m)}(\lambda)=\sum_{i=1}^D A^{(mi)} s^{(i)}(\lambda),\;\; m=1,2,\dots,
\end{eqnarray}
with the diagonal constant matrices $A^{(mi)}$.
 In eq.(\ref{R}),  
$R_0$ is some matrix functions of spectral parameters such that the operator $R(\lambda,\mu)$ is invertible.
We refer to the inverse of $R$  as the spectral function  $\Psi$:
\begin{eqnarray}\label{int}
\int d\Omega(\nu) R(x,\lambda,\nu)\Psi(x;\nu,\mu) \equiv R*\Psi  =I \delta(\lambda-\mu),
\end{eqnarray}
where $\Omega(\nu)$ is some scalar measure on the plane of complex spectral parameter $\nu$, $I$ is the  $N\times N$ 
identity operator.
Differentiating $\Psi$ with respect to $x_m$ and taking into account its definition (\ref{int}) and expression (\ref{R}) 
for $R$  we obtain
\begin{eqnarray}\label{Psix}
&&
\Psi_{x_m}(\lambda,\mu) = -\Psi(\lambda,\mu) s^{(m)}(\mu) +s^{(m)}(\lambda) \Psi(\lambda,\mu)+
\\\nonumber
&&
\Big(\Psi(\lambda,\nu) s^{(m)}(\nu)\Big)* r(\nu) q(\tilde\nu)* \Psi(\tilde\nu,\mu) -
\Psi(\lambda,\nu)* r(\nu) 
q(\tilde\nu)* \Big(s^{(m)}(\tilde\nu)\Psi(\tilde\nu,\mu)\Big),\\\nonumber
&& m=1,\dots,D.
\end{eqnarray}
\begin{eqnarray}\label{Psit}
&&
\Psi_{t_m}(\lambda,\mu) = -\Psi(\lambda,\mu) S^{(m)}(\mu) +S^{(m)}(\lambda) \Psi(\lambda,\mu)+
\\\nonumber
&&
\Big(\Psi(\lambda,\nu) S^{(m)}(\nu)\Big)* r(\nu) q(\tilde\nu)* \Psi(\tilde\nu,\mu) -\Psi(\lambda,\nu)* r(\nu) 
q(\tilde\nu)* \Big(S^{(m)}(\tilde\nu)\Psi(\tilde\nu,\mu)\Big),\\\nonumber
&& m=1,2\dots .
\end{eqnarray}
System (\ref{Psix},\ref{Psit}) can be viewed as a   system for the spectral function $\Psi$.

\subsection{Derivation of system of nonlinear PDEs (\ref{Nwave})}
Now, applying the operators $(q s^{(i)})*$ and $*(s^{(j)} r)$, $i=0,1,2$ (with 
$s^{(0)}=1$) to eqs.(\ref{Psix}) and (\ref{Psit})  we obtain the system of nonlinear PDEs
for the matrix fields
\begin{eqnarray}\label{field1}
&&
v^{(ij)}= (q s^{(i)})*\Psi*(s^{(j)} r), \\\label{field2}
&&
v^{(in0j)}= (q s^{(i)}s^{(n)})*\Psi*(s^{(j)} r),
\;\;\;v^{(i0nj)}= (q s^{(i)})*\Psi*(s^{(n)}s^{(j)} r),\\\nonumber
&&
v^{(i00j)}=v^{(i0j0)}=v^{(0i0j)} \equiv v^{(ij)} .
\end{eqnarray}
This system reads:
\begin{eqnarray}\label{nl1}
&&
v^{(ij)}_{x_m}+v^{(i0mj)} -v^{(im0j)} -
v^{(im)} v^{(0j)} +v^{(i0)} v^{(mj)}=0,\\\label{nl2}
&&
v^{(ij)}_{t_m}+\sum_{k=1}^D \Big(v^{(i0kj)} A^{(mk)}  - A^{(mk)} v^{(ik0j)}\Big)-\\\nonumber
&&
\sum_{k=1}^D v^{(ik)} A^{(mk)} v^{(0j)} +v^{(i0)}\sum_{k=1}^D A^{(mk)} v^{(kj)}=0 ,
\end{eqnarray}
Using the proper combination of eqs.(\ref{nl1},\ref{nl2}) we can eliminate the fields 
$v^{(i0mj)}$ and  $v^{(im0j)}$ writing the system for the fields (\ref{field1}).
For instance, taking $D=2$ we obtain the following nonlinear system (written in components):
\begin{eqnarray}\label{spnlin}
\left|
\begin{array}{ccccc}
(e_{t_1}^{(ij)})_{\alpha\beta} &(e_{t_2}^{(ij)})_{\alpha\beta} &(e_{t_3}^{(ij)})_{\alpha\beta} & (e_{x_1}^{(ij)})_{\alpha\beta} &(e_{x_2}^{(ij)})_{\alpha\beta}\cr
     A^{(11)}_{\beta}  &    A^{(21)}_{\beta}  &        A^{(31)}_{\beta}  &                    1&                      0 \cr
     A^{(11)}_{\alpha} &    A^{(21)}_{\alpha} &        A^{(31)}_{\alpha} &                    1&                      0\cr
     A^{(12)}_{\beta}  &    A^{(22)}_{\beta}  &        A^{(32)}_{\beta}  &                    0&                      1\cr
     A^{(12)}_{\alpha} &    A^{(22)}_{\alpha} &        A^{(32)}_{\alpha} &                    0&                      1
\end{array}
\right|=0
\end{eqnarray}
where we use the following notations:
\begin{eqnarray}
&&
e_{x_m}^{(ij)}:=v^{(ij)}_{x_m}-
v^{(im)} v^{(0j)} +v^{(i0)} v^{(mj)},\\\nonumber
&&e_{t_m}^{(ij)}:=v^{(ij)}_{t_m} -
\sum_{k=1}^2 v^{(ik)} A^{(mk)} v^{(0j)} +v^{(i0)}\sum_{k=1}^2 A^{(mk)} v^{(kj)} ,
\end{eqnarray}
One can see that eq.(\ref{spnlin})  can be written as eq.(\ref{Nwave})
with
\begin{eqnarray}\label{c1}
Z^{1}_{\alpha\beta}=\left|
\begin{array}{cccc}
        A^{(21)}_{\beta}  &        A^{(31)}_{\beta}  &                    1&                      0 \cr
         A^{(21)}_{\alpha} &        A^{(31)}_{\alpha} &                    1&                      0\cr
       A^{(22)}_{\beta}  &        A^{(32)}_{\beta}  &                    0&                      1\cr
         A^{(22)}_{\alpha} &        A^{(32)}_{\alpha} &                    0&                      1
\end{array}
\right|,\;\;
Z^{2}_{\alpha\beta}=-\left|
\begin{array}{cccc}
     A^{(11)}_{\beta}  &         A^{(31)}_{\beta}  &                    1&                      0 \cr
     A^{(11)}_{\alpha} &         A^{(31)}_{\alpha} &                    1&                      0\cr
     A^{(12)}_{\beta}  &            A^{(32)}_{\beta}  &                    0&                      1\cr
     A^{(12)}_{\alpha} &          A^{(32)}_{\alpha} &                    0&                      1
\end{array}
\right|,\;\;
Z^{3}_{\alpha\beta}=\left|
\begin{array}{cccc}
     A^{(11)}_{\beta}  &    A^{(21)}_{\beta}  &                        1&                      0 \cr
     A^{(11)}_{\alpha} &    A^{(21)}_{\alpha} &                        1&                      0\cr
     A^{(12)}_{\beta}  &    A^{(22)}_{\beta}  &                          0&                      1\cr
     A^{(12)}_{\alpha} &    A^{(22)}_{\alpha} &                           0&                      1
\end{array}
\right|,\;\;
\end{eqnarray}
\begin{eqnarray}\label{c2}
T^{1}_{\alpha\beta}=\left|
\begin{array}{cccc}
     A^{(11)}_{\beta}  &    A^{(21)}_{\beta}  &        A^{(31)}_{\beta}  &                                         0 \cr
     A^{(11)}_{\alpha} &    A^{(21)}_{\alpha} &        A^{(31)}_{\alpha} &                                         0\cr
     A^{(12)}_{\beta}  &    A^{(22)}_{\beta}  &        A^{(32)}_{\beta}  &                                        1\cr
     A^{(12)}_{\alpha} &    A^{(22)}_{\alpha} &        A^{(32)}_{\alpha} &                                          1
\end{array}
\right|,\;\;
T^{2}_{\alpha\beta}=\left|
\begin{array}{cccc}
     A^{(11)}_{\beta}  &    A^{(21)}_{\beta}  &        A^{(31)}_{\beta}  &                    1 \cr
     A^{(11)}_{\alpha} &    A^{(21)}_{\alpha} &        A^{(31)}_{\alpha} &                    1\cr
     A^{(12)}_{\beta}  &    A^{(22)}_{\beta}  &        A^{(32)}_{\beta}  &                    0\cr
     A^{(12)}_{\alpha} &    A^{(22)}_{\alpha} &        A^{(32)}_{\alpha} &                    0
\end{array}
\right|.
\end{eqnarray}
\begin{eqnarray}\label{c3}
a^{(k)}_{\alpha\gamma\beta} =\left|
\begin{array}{ccccc}
A^{(1k)}_\gamma  &A^{(2k)}_\gamma  &A^{(3k)}_\gamma  & \delta_{k1} &\delta_{k2} \cr
     A^{(11)}_{\beta}  &    A^{(21)}_{\beta}  &        A^{(31)}_{\beta}  &                    1&                      0 \cr
     A^{(11)}_{\alpha} &    A^{(21)}_{\alpha} &        A^{(31)}_{\alpha} &                    1&                      0\cr
     A^{(12)}_{\beta}  &    A^{(22)}_{\beta}  &        A^{(32)}_{\beta}  &                    0&                      1\cr
     A^{(12)}_{\alpha} &    A^{(22)}_{\alpha} &        A^{(32)}_{\alpha} &                    0&                      1
\end{array}
\right|,\;\;k=1,2.
\end{eqnarray}
The structure of eq.(\ref{Nwave}) is similar to that of  (2+1)-dimensional $n$-wave equation (\ref{Nw}). 
In particular, it is off-diagonal and admits the  reduction
\begin{eqnarray}\label{red}
&&
x_n= i \xi_n,\;\;t_m= i \tau_m,\;\;,\;\;
(v^{(lj)})^+=v^{(jl)},\\\nonumber
&&
i^2=-1,\;\;n=1,2,\;\;m=1,2,3,\;\;l,j=0,1,2.
\end{eqnarray}
Under this reduction,  system (\ref{Nwave}) reads
\begin{eqnarray}\label{NwaveRed}
&&
i\sum_{m=1}^3 Z^{(m)}_{\alpha\beta}(v^{(lj)}_{\alpha\beta})_{\tau_m}+
i  \sum_{n=1}^2 T^{(n)}_{\alpha\beta}(v^{(lj)}_{\alpha\beta})_{\xi_n} + \sum_{k=1}^2
\sum_{\gamma=1}^N a^{(k)}_{\alpha\gamma\beta} (v^{(lk)}_{\alpha\gamma} v^{(0j)}_{\gamma\beta} -
 v^{(l0)}_{\alpha\gamma} v^{(kj)}_{\gamma\beta}) =0,\\\nonumber
&&
l,j=0,1,2,\;\;\alpha,\beta=1,\dots,N.
\end{eqnarray}

\section{Supplemented algebraic-differential relations among functions $v^{(ij)}$}
\label{Section:relations}
According to Sec.\ref{Section:dressing}, the off-diagonal elements of the matrix  
fields $v^{(ij)}$, $i,j=0,1,2$ satisfy  system (\ref{Nwave}). 
In this section we show that there are natural supplemented 
algebraic-differential relations among these matrix fields, 
involving  their diagonal 
elements. There are three families of such relations.

1. {\it The first family} is represented by 
the equations of  systems (\ref{nl1},\ref{nl2}) with $i=j=0$:
\begin{eqnarray}\label{r1}
&&
v^{(00)}_{x_m}+v^{(0m)}- v^{(m0)} -
v^{(0m)} v^{(00)} +v^{(00)} v^{(m0)}=0,\\\nonumber
&&
v^{(00)}_{t_m}+\sum_{k=1}^2 (v^{(0k)} A^{(mk)} -  A^{(mk)} v^{(k0)})  -
\sum_{k=1}^2 v^{(0k)} A^{(mk)} v^{(00)} +v^{(00)}\sum_{k=1}^2 A^{(mk)} v^{(k0)} =0 .
\end{eqnarray}

2. {\it The second  family} is  represented by the equations of 
systems (\ref{nl1},\ref{nl2}) with $j=0$.  Eliminating the fields  $v^{(im00)}$ 
we write this family as 
\begin{eqnarray}\label{r2}
\left|
\begin{array}{ccccc}
(f_{t_k}^{(i0)})_{\alpha\beta}  & (f_{x_1}^{(i0)})_{\alpha\beta} &(f_{x_2}^{(i0)})_{\alpha\beta}\cr
     A^{(k1)}_{\alpha} &                     1&                      0\cr
     A^{(k2)}_{\alpha} &                       0&                      1
\end{array}
\right|=0,\;\;k=1,2,3,
\end{eqnarray}
\begin{eqnarray}\nonumber
&&
f_{x_m}^{(i0)}:=v^{(i0)}_{x_m} +v^{(im)} -
v^{(im)} v^{(00)} +v^{(i0)} v^{(m0)} ,\\\nonumber
&&f_{t_m}^{(i0)}:=v^{(i0)}_{t_m}+\sum_{k=1}^2 v^{(ik)} A^{(mk)} -
\sum_{k=1}^2 v^{(ik)} A^{(mk)} v^{(00)} +v^{(i0)}\sum_{k=1}^2 A^{(mk)} v^{(k0)} ,
\end{eqnarray}

3.{\it The third  family} is  represented by the equations of 
systems (\ref{nl1},\ref{nl2}) with $i=0$.  Eliminating the fields $v^{(00mj)}$  
we write this family as 
\begin{eqnarray}\label{r3}
\left|
\begin{array}{ccccc}
(f_{t_k}^{(0j)})_{\alpha\beta}  & (f_{x_1}^{(0j)})_{\alpha\beta} &(f_{x_2}^{(0j)})_{\alpha\beta}\cr
     A^{(k1)}_{\beta}  &                    1&                      0 \cr
     A^{(k2)}_{\beta}  &                    0&                      1\cr
\end{array}
\right|=0,\;\;k=1,2,3,
\end{eqnarray}
\begin{eqnarray}\nonumber
&&
f_{x_m}^{(0j)}:=v^{(0j)}_{x_m} -v^{(mj)} -
v^{(0m)} v^{(0j)} +v^{(00)} v^{(mj)} ,\\\nonumber
&&
f_{t_m}^{(0j)}:=v^{(0j)}_{t_m}-\sum_{k=1}^2  A^{(mk)} v^{(kj)}-
\sum_{k=1}^2 v^{(0k)} A^{(mk)} v^{(0j)} +v^{(00)}\sum_{k=1}^2 A^{(mk)} v^{(kj)} ,
\end{eqnarray}

All relations (\ref{r1}-\ref{r3}) have the diagonal 
parts and involve diagonal elements of all matrix fields $v^{(ij)}$, $i,j=0,1,2$. 
However, we have to emphasize that the presence of these relations does not reduce the 
complete integrability of the system  (\ref{Nwave}). This statement will be clarified in
Sec.\ref{Section:solutions} 

\section{Solution space of system  (\ref{Nwave})}
\label{Section:solutions}
First of all, in this section we show the complete integrability of  system (\ref{Nwave}).
For this purpose we again turn to the spectral representation 
and derive the explicite form of the spectral function $\Psi$ and matrix fields $v^{(ij)}$.

\subsection{Derivation of explicite formulas for $\Psi$ and $v^{(ij)}$}
\label{Section:explicite}
To construct the explicite form of $\Psi$ we have to invert the operator $R$,
i.e., to solve the integral equation (\ref{int}) for  $\Psi$. It is remarkable that $\Psi$ can be constructed explicitly 
up to the invertibility of the integral operator $R_0$ independent on $x$- and $t$-variables.
Substituting $R$ from (\ref{R}) we rewrite eq.(\ref{int}) as  
\begin{eqnarray}
\int \hat R_0(\lambda,\nu) \Psi(\nu,\mu) d\Omega(\nu) + r(\lambda) \chi(\mu) = \delta(\lambda-\mu),
\end{eqnarray}
where
\begin{eqnarray}\label{R0}
&&
\hat R_0(\lambda,\nu) =e^{ \sum_{j=1}^D s^{(j)}(\lambda) x_j +\sum_{k} S^{(k)}(\lambda) t_k } 
R_0(\lambda,\mu)  e^{ -\sum_{j=1}^D  s^{(j)}(\mu) x_j - \sum_{k} S^{(k)}(\lambda) t_k} ,\;\;
\\\label{chi}
&&
\chi(\mu)= \int q(\nu) \Psi(\nu,\mu) d\Omega(\nu).
\end{eqnarray}
Assuming the invertibility of $R_0(\lambda,\mu)$, we can write
\begin{eqnarray}
\hat R_0^{-1}(\lambda,\mu)= e^{ \sum_{j=1}^D  s^{(j)}(\lambda) x_j + 
\sum_{k} S^{(k)}(\lambda) t_k} R_0^{-1}(\lambda,\mu) 
e^{ -\sum_{j=1}^D s^{(j)}(\mu) x_j -\sum_{k} S^{(k)}(\mu) t_k }, 
\end{eqnarray}
where  $R_0(\lambda,\nu)*R_0^{-1}(\nu,\mu)=\delta(\lambda-\mu)$.
Then, applying $\hat R_0^{-1}*$, we can find $\Psi$ as 
\begin{eqnarray}\label{Psi}
\Psi=\hat R_0^{-1} - \hat R_0^{-1}* r \chi.
\end{eqnarray}
The function $\chi$ in the rhs of this equation can be found applying  $q*$ to eq.(\ref{Psi}) and solving the 
obtained equation for $\chi$:
\begin{eqnarray}
\chi = (1+w^{(00)}) q*\hat R_0^{-1},
\end{eqnarray}
where we introduce the following  notations:
\begin{eqnarray}\label{w}
w^{(ij)} = (qs^{(i)})*\hat R_0^{-1}*(s^{(j)}r),\;\;i,j=0,1,\dots,D,\;\; s^{(0)}=I_N.
\end{eqnarray}
Applying $(qs^{(i)})*$ and $*(s^{(j)}r)$ to eq.(\ref{Psi}) we obtain:
\begin{eqnarray}\label{v}
v^{(ij)} = w^{(ij)} - w^{(i0)} (I + w^{(00)})^{-1} w^{(0j)}.
\end{eqnarray}
Obviously, each of the functions $w^{(ij)}$ is a solution of the appropriate linear PDE
(\ref{Nwave_lin}).
Solutions $v^{(ij)}$ have no singularities if ${\mbox{det}}(I + w^{(00)})\neq 0$ 
for all $x_i$ and $t_i$ inside of 
the domain of our interest.

The conclusion  regarding the complete integrability of  system (\ref{Nwave}) 
follows from the fact that the function $R$ (\ref{R0}) has arbitrary dependence on $2 D$ independent 
 functions of spectral parameters $s^{(i)}(\lambda)$ and $s^{(i)}(\mu)$  
thus providing arbitrary functions of $2 D$ independent variables $x_i$ and $t_i$. 
This is a necessary requirement for complete integrability of the system of 
$(2 D+1)$-dimensional PDEs. Obviously, the number of these arbitrary matrix functions of $(x,t)$-variables 
coincides 
with the number of fields $v^{(ij)}$ 
in  system (\ref{Nwave}). In fact, these arbitrary functions are nothing but the fields 
$w^{(ij)}$ (\ref{w}). By construction, 
they are independent functions having arbitrary dependence on $2 D$ variables $x_i$ and $t_i$. 

In addition, to  satisfy the reduction (\ref{red}), we shall impose the 
following constraints on the functions of spectral parameters 
$R_0$ and $s$:
\begin{eqnarray}\label{redRrqs}
R_0(\lambda,\mu) = R_0^+(\mu^*,\lambda^*),
\;\;q(\lambda)=r^*(\lambda^*),
\;\;s^{(n)}(\lambda) = (s^{(n)})^*(\lambda^*),
\end{eqnarray}
where the star means the complex conjugate and we take into account the diagonality of $r$, $q$ and $s^{(n)}$.

\subsection{Particular solutions }

\begin{figure*}
   \epsfig{file=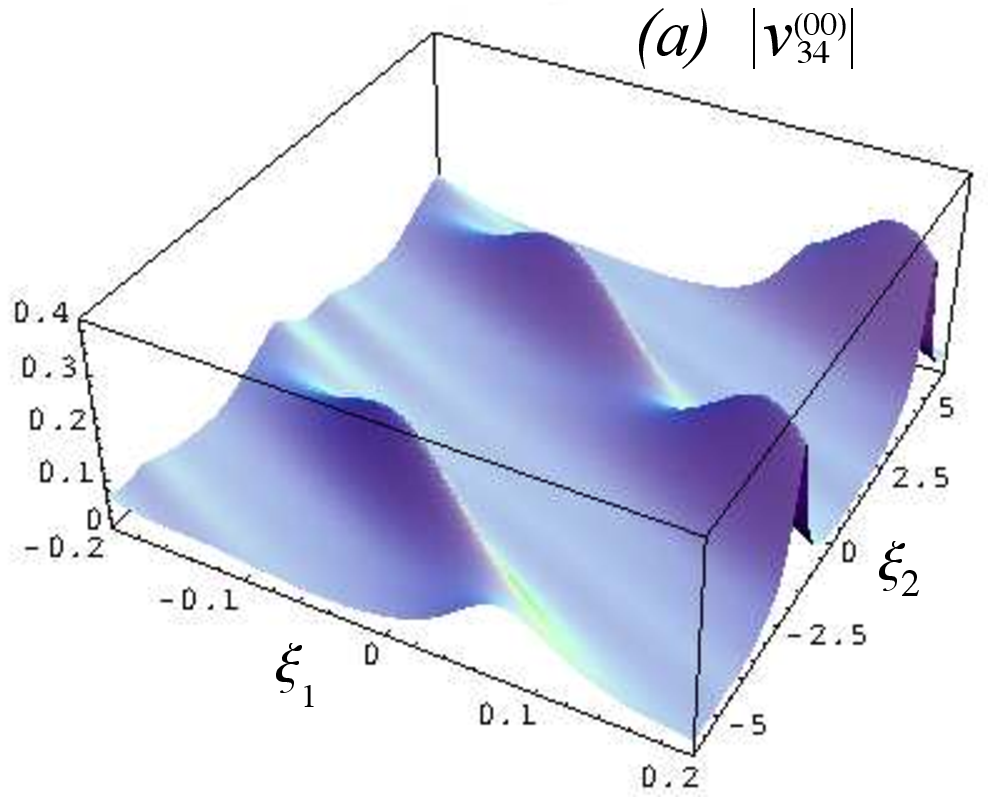,
  scale=0.8
   ,angle=0
} \epsfig{file=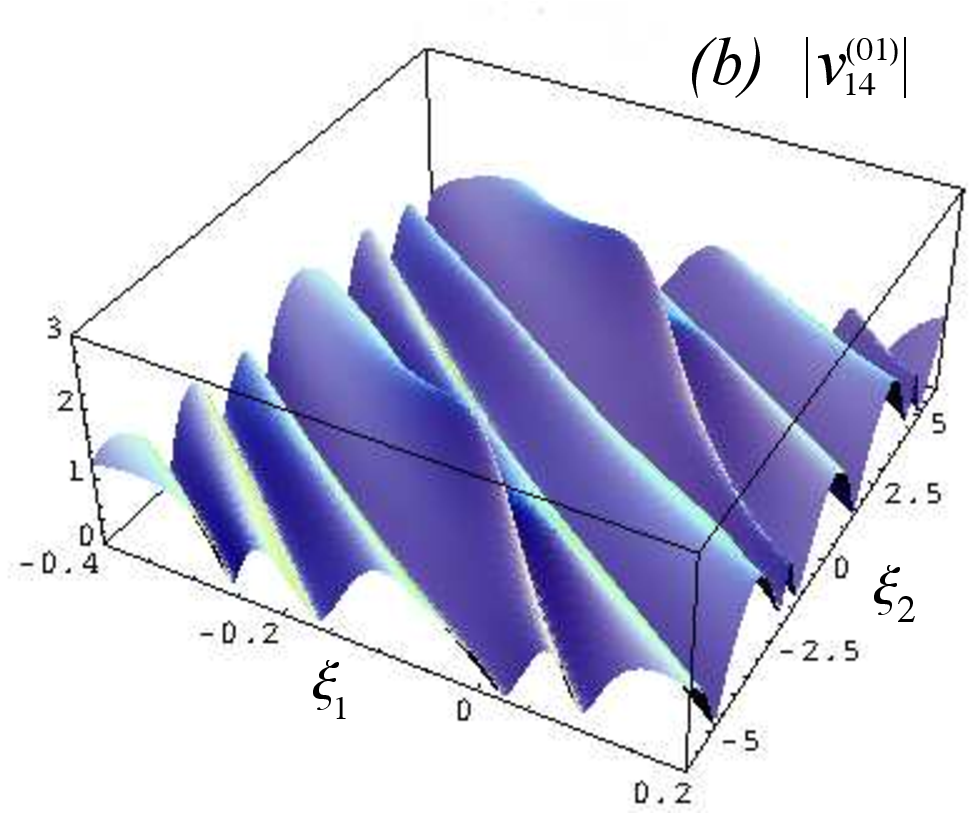,
  scale=0.8
   ,angle=0
}\\
\epsfig{file=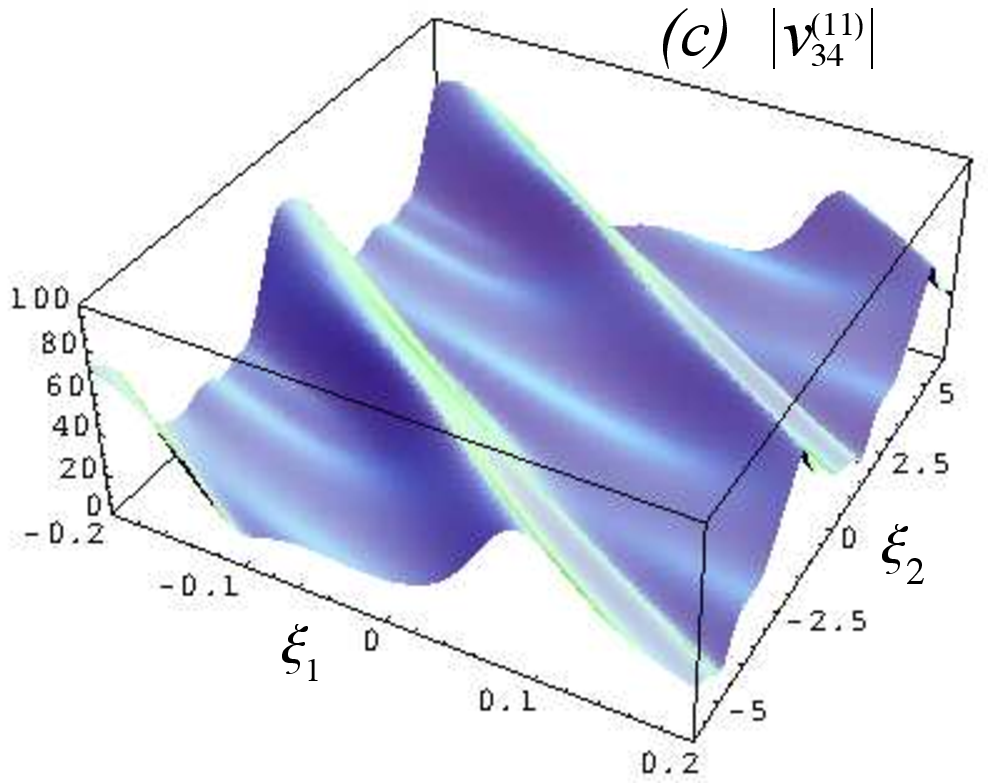,
  scale=0.8
   ,angle=0
}

\caption{The absolute values of some matrix elements of the  particular solution
to eq.(\ref{NwaveRed}) at  $\tau_i=i$, $i=1,2,3$. 
Three  typical cases  are represented.
(a) The absolute value of the matrix  elements of $v^{(00)}$ and $v^{(10)}$ have the 
set of  upward peaks distributed  on the $(\xi_1,\xi_2)$-plane,  $|v^{(00)}_{34}|$ is shown as an example. 
(b)  The absolute value of the matrix  elements of $v^{(01)}$ have upward peaks distributed over the oscillating basis, 
$|v^{(01)}_{14}|$  is shown as an example.
(c) The absolute value of the matrix elements of $v^{(11)}$ have oscillating form without sharp peaks, 
$|v^{(11)}_{34}|$  is shown as an example.
} 
  \label{Fig:sol} 
\end{figure*}

In this section we consider the particular solutions $v^{(ij)}$ under the 
reduction (\ref{red}) (i.e., solutions of eq. (\ref{NwaveRed})) constructed 
using the discrete spectral parameters: 
$\lambda,\mu\in \{z_1,z_2,\dots, z_M\}$ with real $z_i$. 
As an example we consider the following functions $R_0$, $r$, $q$ and $s^{(n)}$ satisfying 
reduction (\ref{redRrqs}):
\begin{eqnarray}
&&
(R_0(z_j, z_k))_{\alpha\beta}=
\delta_{\alpha\beta}\delta_{jk} + \alpha^{z_j} \beta^{z_k} (1-\delta_{jk})
,\\\nonumber
&&
s^{(n)}_\alpha(z_j) =  (\alpha z_j)^n,\\\nonumber
&&
A^{(mn)}_\alpha=(\alpha+n-1)^{m},\\\nonumber
&&
r_\alpha(z_j)=q_\alpha(z_j)= \alpha^{z_j},\\\nonumber
&&
n=1,2,\;\;m=1,2,3,\;\;j,k=1,\dots,M,\;\;\alpha,\beta=1,\dots,N,
\end{eqnarray}
{
where $\delta_{ij}$ is the Kronecker symbol.
In this case, the fields $v^{(ij)}$ are given by eq.(\ref{v})  
with $w^{(ij)}$ defined by  eqs.(\ref{w}) and (\ref{R0}):
\begin{eqnarray}\label{wd}
w^{(ij)} = \sum_{i,j=1}^M \alpha^{z_i} (\alpha z_i)^n 
e^{K_\alpha(x,t,z_i)-K_\beta(x,t,z_j) } ( R_0^{-1})_{\alpha\beta}(z_i,z_j)  (\beta z_j)^m \beta^{z_j}
\end{eqnarray}
where
\begin{eqnarray}
&&
K_\alpha(x,t,z_i) = \sum_{j=1}^2 (\alpha z_i)^j x_j + 
\sum_{l=1}^3\sum_{k=1}^2 (\alpha + k -1)^l (\alpha z_i)^k t_l,
\end{eqnarray}
and $R_0^{-1}$ is the inverse of $R_0$:
\begin{eqnarray}
&&
\sum_{l=1}^M \sum_{\gamma=1}^N \Big(\delta_{\alpha\gamma} \delta_{il} + 
\alpha^{z_i} \gamma^{z_l}(1-\delta_{il})\Big) ( R_0^{-1})_{\gamma\beta}(z_l,z_j) =
 \delta_{\alpha\beta}\delta_{ij}.
\end{eqnarray}
}
The absolute values of some matrix elements of $v^{(ij)}$  for the case $M=2$ and $N=4$ are depicted in Fig.\ref{Fig:sol} 
on the plane $(\xi_1,\xi_2)$ at fixed $\tau_i=i$.
We see that all matrix  elements of $v^{(ij)}$ can be separated into three family. 
The absolute values of matrix  elements from the  first family (of the elements of $v^{(00)}$ and $v^{(10)}$, 
Fig.\ref{Fig:sol}a) have high peaks 
regularly situated on the $(\xi_1,\xi_2)$-plane. The absolute values of the matrix  elements from the second family (of 
the elements of 
$v^{(01)}$, Fig.\ref{Fig:sol}b)
have  peaks situated on the oscillating basis. 
Finally,  the absolute values of the matrix elements from the  third family 
 (of the elements of $v^{(11)}$, Fig.\ref{Fig:sol}c)
have no sharp peaks. So, such $n$-wave type equation can be responsible for forming the 
lump-type quasi-periodic solutions of large amplitude.

 \section{Conclusions}
 \label{Section:conclusion}
 We construct a completely integrable linearizable $5$-dimensional  version of  $n$-wave 
 type equation for the matrix fields $v^{(ij)}$,
 $i,j=0,1,2$. Its generalization to $2 D +1$, $D>2$, case is obvious. 
  We show that the elements of the matrix fields $v^{(ij)}$ have different shapes. Some of them 
  can be viewed as a set of quasi-periodic sharp peaks and thus could be responsible for formation of 
  rogue waves. 
 
 It is remarkable, 
 that the auxiliary spectral function $\Psi$ was introduced 
 to  derive eq.(\ref{Nwave}), so that fields $v^{(ij)}$ acquire the spectral representation, 
 similar to the fields of ISTM-integrable 
 nonlinear PDEs. Thus, the dressing method appears here as a tool unifying the  
 nonlinear PDEs integrable by  different methods. 
 Similar phenomena have been observed in
 earlier papers \cite{Z_2004,Z_JMP_2009}, combining nonlinear PDEs integrable 
 by different methods, such as ISTM, 
 linearization by the Hopf substitution, method of characteristics.

 Of course, instead of the symmetric form 
 of $\hat R_0$ (\ref{R0}) we could take its non-symmetric version
 
 \begin{eqnarray}\label{R0nonsym}
&&
\hat R_0(\lambda,\nu) =e^{ \sum_{j=1}^D s^{(j)}(\lambda) x_j +\sum_{k} S^{(k)}(\lambda) t_k } 
R_0(\lambda,\mu)  e^{ \sum_{j=1}^D  p^{(j)}(\mu) x_j + \sum_{k} P^{(k)}(\lambda) t_k}.
\end{eqnarray}
Then the  appropriate nonlinear PDEs  would involve the diagonal parts as well.

 The work is partially supported by  the Program for Support of Leading Scientific Schools (grant No. 9697.2016.2), 
 and by the RFBR (grant No. 14-01-00389).



\begin{thebibliography}{99}

\bibitem{K1}
 D. J. Kaup, Stud. Appl. Math. {\bf 62}, 75 (1980).
\bibitem{K2}
 D. J. Kaup, Phys. D {\bf 1}, 45 (1980).

\bibitem{GGKM}
C.S.Gardner, J.M.Green, M.D.Kruskal, R.M.Miura, Phys.Rev.Lett, {\bf 19}, (1967)
1095

\bibitem{ZMNP}
V. E. Zakharov, S. V. Manakov, S. P. Novikov, and L. P. Pitaevsky, 
Theory of Solitons. The Inverse Problem Method (Plenum
Press, 1984).
\bibitem{AC}
M. J. Ablowitz and P. C. Clarkson, Solitons, Nonlinear Evolution Equations and Inverse Scattering (Cambridge University
Press, Cambridge, 1991).

\bibitem{Sh}
A.B.Shabat, Dokl. Akad. Nauk SSSR {\bf 211} No. 6, (1973) 1310


\bibitem{ZSh1}
V.E.Zakharov and A.B.Shabat, Funk.Anal.Pril., {\bf 8}, (1974) 43;
Funct.Anal.Appl., {\bf 8}, (1974) 226

\bibitem{ZSh2}
V.E.Zakharov and A.B.Shabat, Funct.Anal.Appl. {\bf 13}, (1979) 13;
Funk.Anal.Pril., {\bf 13}, (1979) 166


\bibitem{ZM}
V.E.Zakharov and S.V.Manakov, Funk.Anal.Pril. {\bf 19}, (1985) 11;
Funct.Anal.Appl. {\bf 19}, (1985) 89

\bibitem{K}
 B. Konopelchenko, Solitons in Multidimensions (World Scientific, Singapore, 1993).

\bibitem{Z_JMP_2014}
A.I.Zenchuk, J.Math.Phys. {\bf 55}, 121505 (2014)

 
\bibitem{Calogero}
F. Calogero in {\it What is Integrability} ed V.E.Zakharov (Berlin: Springer)
(1990) 1


\bibitem{CX1}
F. Calogero and Ji Xiaoda, J. Math. Phys. {\bf 32}, 875 (1991).

\bibitem{CX2}
F. Calogero and Ji Xiaoda, J. Math. Phys. {\bf 32}, 2703 (1991).

 
 \bibitem{Z_2004}
A. I. Zenchuk, J. Phys. A: Math.Gen. {\bf 37}, 6557 (2004).

 
 
 \bibitem{Z_JMP_2009}
 A. I. Zenchuk, 
J.Math.Phys. {\bf 50}, 063505 (2009)


\end{thebibliography}
\end{document}